\documentclass[11pt]{article}
\usepackage{pst-all}
\usepackage{pstricks}
\usepackage{pstcol,pst-fill,pst-grad}
\usepackage{amsfonts}
\usepackage{graphicx}
\usepackage{amsmath}
\usepackage[normalem]{ulem}
\usepackage{multicol}
\usepackage{tikz}
\usetikzlibrary{matrix}
\usepackage{enumerate}
\RequirePackage{mathrsfs} \RequirePackage[sc]{mathpazo}
\RequirePackage{wasysym} \RequirePackage{setspace}

\makeatletter \@addtoreset{equation}{section}

\newcommand{\be}{\begin{equation}}
\newcommand{\ee}{\end{equation}}
\newcommand{\bea}{\begin{eqnarray}}
\newcommand{\eea}{\end{eqnarray}}
\begin{document}
\date{}
\title{  Weighted Graph  Theory  Representation  of  Quantum  Information  Inspired by   Lie Algebras}
\author{ Abdelilah Belhaj$^{1}$, Adil Belhaj$^{2}$, Larbi
Machkouri$^{3}$, Moulay Brahim  Sedra$^{3}$, Soumia Ziti$^{1}$
\hspace*{-8pt} \\
%EndAName
\\
 {\small $^{1}$ Labortoire de Recherche en Informatique, Facult\'{e}
des Sciences,  Universit\'e Mohammed V,
}\\{ \small Rabat, Morocco}\\
 {\small $^{2}$ LIRST, D\'epartement de Physique, Facult\'e
Polydisciplinaire, Universit\'e Sultan Moulay Slimane}\\{ \small
B\'eni Mellal, Morocco }
\\ {\small $^{3}$   LabSIMO,   D\'{e}partement de Physique, Facult\'{e}
des Sciences, Universit\'{e} Ibn Tofail }\\{ \small K\'{e}nitra,
Morocco} }\maketitle

\begin{abstract}
 Borrowing ideas from the relation between  simply laced Lie
algebras and  Dynkin diagrams,
 a weighted  graph theory representation of quantum  information is addressed.  In this way,
  the density matrix of a quantum  state can be  interpreted as
 a signless Laplacian matrix of an associated  graph. Using  similarities with  root
systems of simply laced Lie algebras,  one-qubit theory is analyzed
in some details and is found to be linked to  a  non-oriented
weighted graph having two vertices. Moreover, this  one-qubit theory
is generalized to $n$-qubits. In this representation, quantum gates
correspond to graph weight operations preserving the probability
condition. A speculation from string theory, via D-brane quivers, is
also given.
\newline
\newline{\bf Keys words }:  Graph Theory;  Laplacian Matrix;  Density
Matrix; Quantum Gates;   Dynkin Diagram; Cartan Matrix.
\end{abstract}

%\newpage
%\tableofcontents

\thispagestyle{empty}

\newpage \setcounter{page}{1} \newpage
\section{Introduction}

Quantum information   combines computer science and quantum
mechanics. It has been remarked that  such a combination can be
explored to produce computers operating according to quantum
mechanics providing fast calculations and simulations using the
properties of quantum bits. For these reasons,  quantum information
theory has been extensively investigated in connections with quantum
algorithms and communication protocols \cite{1,2}. This theory is
based on quantum systems  approached using mathematical backgrounds
dealing with density matrices and operators associated with
tensor-product of Hilbert  vector spaces.  These fundamental pieces
are    quite relevant in the discussion of interesting quantum
phenomenons including entanglement\cite{3,4}.

 Recently,
qubit systems  have been  studied  using different approaches
including  string theory and related models \cite{5,6,7,70}. These
investigations  have brought new understanding of the fundamental
physics associated with  qubits and theirs supersymmetric
extensions. The latters are connected to many theories including
D-branes, toric geometry and supermanifolds. More precisely, a nice
interplay between the black holes and qubits have been discussed
using  higher dimensional supergravity models\cite{5,6,8}.
Alternative studies have been conducted using  toric geometry and
Adinkra graph theory\cite{9,10}. This graph  theory has been used in
the study of the supersymmetric representation  of quantum field
theories. The corresponding graphs are formed by nodes associated
with bosonic and fermionic
 degree of freedom. These   graphs have been used to classify a class of
 qubit  black  holes \cite{11}.  Alternatively,  a graphic representation  has
been also explored to approach  quantum states  using   Seidel
switching for weighted multidigraphs \cite{12}.

 The aim of this work  is to contribute to these activities by considering
 a weighted graph theory to  represent basic pieces of quantum  information.
The present proposition  can bring  more  features on  quantum
information operations. In fact, unitary evolutions will be
displayed by modifications over their graph representations. Instead
of performing  quantum computations, the proposed graph
representation allows one to encode qubit physical properties in
terms of simple combinatorial data. This realization, which has
similarities with quivers  used in string theory,  may permit to
extract the essential on the qubit physics by simply knowing graph
vertices and edges. Inspired by the relation between  simply laced
Lie algebras and Dynkin diagrams, we first interpret the
corresponding density matrices  as signless Laplacian matrices  of
weighted graphs. Borrowing ideas from  root  systems,  we discuss
one-qubit theory in some details and show that it is linked to  two
weighted vertices sharing  similarities with $A_2$ quivers placed on
singularities associated with   D-brane physics.  This one-qubit
theory is generalized to $n$-qubits. Then, we reveal that the
quantum gates  correspond to graph  weight operations preserving the
probability condition. We finish with speculations motivated by
string theory and quivers.

The organization of this paper is as follows. In section 2, we
present a weighted graph theory of  one-qubit systems.  Section 3
concerns a general discussion involving higher dimensional qubits.
Operations on graph weights  are explored  in Section 4 to discuss
universal quantum gates. Section 5 is devoted to open questions and
speculation supported by string theory.

\section{Graph  theory representation of quantum states}

In this section, we establish a rigorous correspondence between
quantum states and  weighted graph theory. The present approach is
firstly inspired by Lie algebra structure and partly by the work
given in \cite{12} in which a quantum state is represented by its
corresponding and identifying graph. In some related works, every
vertex graph is considered to be a quantum state with the edge being
the interaction between vertices. However, in the present paper, a
quantum system is represented by a graph that is associated with
 a single element in the  Hilbert space. The bridge employed
here is the density matrix of a quantum state that is identified
with the Signless Laplacian matrix   of  the corresponding graph.

\subsection{Graph theory basics}
In this subsection, we give a concise review on graph theory. More
details  can be found in  \cite{13,14,15}. Indeed, a  graph is
mathematically defined by a pair of sets $G = (V(G), E(G))$, where
$V(G)$ denotes the vertex set and $E(G)$  corresponds the edge set.
Two vertices are said adjacent if they are connected by an edge. For
instance, if the vertices $i$ and $j$ are linked, the edge is
indexed by $(i,j)$. The number of edges adjacent to a vertex is
called its
degree,  and it is denoted by $d_{G}(v_{i})$  where $v_{i}$ represents the vertex indexed by $i$.\\
To a graph $G$ we associate a symmetric matrix called an adjacency
matrix $M(G) = (m_{ij})$, whose elements are either $0$ or $1$. The
adjacency matrix  which  provides  connections between different
vertices  encodes  all the information residing on the graph. These
two different objects share the same  data which can explored either
in mathematics  or physics including string  theory and related
quiver  models. Roughly,  the adjacency  matrix   is defined as
follows
\begin{equation}
m_{ij}=\left\{%
\begin{array}{ll}
    1, &   (i,j)\in E(G), \\
    0, &   (i,j)\not\in E(G). \\
\end{array}%
\right.
\end{equation}

It is recalled that  the adjacency matrix $A(G)$ order is the
cardinal of the vertices set $V(G)$ being the  number of its
elements. Besides the adjacency matrix, we also associate to a graph
a diagonal matrix of the same order  called the degree matrix
$D(G)$. Its $i$-th diagonal element is the degree of the $i$-th
vertex of $G$, with  $i = 1, 2, \ldots, |V|$,  where $|V|$ is the
cardinal of the set $V(G)$. In fact, it is given by
\begin{equation}
d_{ij}=\left\{%
\begin{array}{ll}
    d_{G}(v_{i}), & \hbox{ if} \;\; i = j, \\
    0 , & \hbox{  if}  \;\; i \neq j. \\
\end{array}%
\right.
\end{equation}
In graph theory, one can define the Laplacian matrix $L(G)$ also
called the admittance matrix  by combining  the degree matrix $D$
and the adjacency matrix $M$. It  is defined as follows
\begin{equation}
L(G) = D(G) - M(G).
\end{equation}
In such a  literature,  we encounter various formulation of
Laplacian matrix. For later use, we consider only the signless
Laplacian matrix
\begin{equation}
Q(G) = D(G) + M(G).
\end{equation}
This matrix will be relevant in the forthcoming sections. More
precisely, it will play a bridge between quantum information pieces
and weighted  graph theory.

\subsection{Weighted graph of the density matrix}

The standard formulation of quantum mechanics is based on the
Hilbert space structures. Indeed, a quantum system state is an
element of a $r$-dimensional Hilbert space
$\mathcal{H}\cong\mathbb{C}^{r}$. For such a state, one defines   an
$r\times r$  positive semidefinite, trace-one, hermitian matrix,
called density matrix. The latter  gives a  general description of a
quantum state which will be represented by a  graph  encoding the
corresponding  quantum  information.

 Motivated by simply laced Lie algebras and Cartan matrices, the emphasis  is  put on the signless Laplacian matrix $Q(G)$ which is
positive, semidefinite and Hermitian matrix. It will be scaled by
$tr(Q(G))$ being defined as  follows
\begin{equation}\rho= \rho_Q(G) = \frac{Q(G)}{tr(Q(G))}.\end{equation}

\subsection{Weighted graph theory representation of one-qubit}

As claimed above,  the description of a quantum state can be made
via  the  density matrix. The corresponding data will be encoded in
a graph.  Before giving the general statement,  we  consider
examples, through which we will illustrate how practically things
work.
 Indeed, there is no simple example to start with than a one-qubit
quantum system. This fundamental piece will play a crucial role in
the coming sections and in the graph operations that we will
establish later on to deal with quantum gates. This simple example
can be considered as a building block for higher dimensional qubit
quantum systems. To start, we illustrate this connection  with a two
vertex model. Then, one can think of many simple examples involving
combinatorial numbers  of vertices to build graphs associated with
multi-qubits.

It is recalled that a one-qubit is a state of a two dimensional
Hilbert space. Employing Dirac notation  associated with the  basis
\{{$|0>$, $|1>$}\}, this state can be written, ($i=\{0,1\}$), as
follows
\begin{equation}|\psi> = a_{0}|0> + a_{1}|1> = \sum_{i=0}^1a_{i}|i>\end{equation}
where $a_{i}$ are complex numbers verifying the normalization
condition \begin{equation}|a_{0}|^2 + |a_{1}|^2= 1.\end{equation}
The coefficient $|a_{i}|^{2}$ corresponds to the probability of
measuring the qubit in the state $|i>$.

Moreover,  the state  $|\psi>$,   used  to describe  one-qubit
quantum systems, can be   associated with  a $2\times2$ matrix,
called the density matrix defined by the following relation
\begin{equation}\rho = |\psi><\psi| = \left(%
\begin{array}{cc}
  a_{0}\overline{a_{0}} & a_{0}\overline{a_{1}} \\
  a_{1}\overline{a_{0}} & a_{1}\overline{a_{1}}\\
\end{array}
\right).\end{equation}
 Motivated  by  physical applications including connection with black holes \cite{5,6,7,8,9}, we will consider real density matrices. In this way,
  they  can be written  generally as follows
\begin{equation}\rho_{ij}=a_{i}a_{j}.\end{equation}
 A close inspection shows that this  equation can be  handled   to provide a
 bridge with graph theory using similarities with Dynkin diagrams of simply laced algebras.   To establish such a link,  it is useful
 to recall Lie algebra theory \cite{16,17}. Indeed,  a  Lie algebra is a vector space $g$  with a bilinear map
  $[,] : g \times g \to g$  (i.e.\ a linear map $[,]: g \otimes g \to g$)
  satisfying the following properties
\begin{enumerate}
  \item antisymmetry: $[x,y] + [y,x] =
  0,$
  \item Jacobi identity: $[x,[y,z]] + [y,[z,x]] + [z,[x,y]] =
  0.$
\end{enumerate}
A  fundamental piece  is the Cartan subalgebra $H$ defined as  the
maximal abelian Lie sub-algebra useful to  decompose  the root space
as follows
    \begin{eqnarray}
g &=&g_{0}\oplus \left\{ \oplus _{\alpha \neq 0}g_{\alpha }\right\}
\nonumber
\\
g &=&H\oplus \left\{ \oplus _{\alpha \neq 0}g_{\alpha }\right\}.
\end{eqnarray}
where $ g_{\alpha}  =\left\{ x\in g, \;\;\left[ h,x\right]
=\alpha(h)x, \;\;\forall h\in H\right\}$. In this way,   the vectors
$\alpha$ are called roots. It is noted that
 a  root system  {{$\Delta$}} of a Lie symmetry is defined
as a subset of an Euclidean space $E$  satisfying the following
constraints
\begin{enumerate}
\item $\Delta$  is finite and spans $E$, $0\notin \Delta,
$
\item if  $ \alpha$ is an element of  $\Delta $, $k\alpha$ is also   if $%
k=\pm 1,$
\item  for all $\alpha$ inside $ \Delta $,  $\Delta$ is invariant under reflections  $\sigma
_\alpha$,
\item  if  $\alpha $  and $\beta $
inside $\Delta$, the quantity $ \left\langle \beta .\alpha
\right\rangle =\frac{2(\beta .\alpha )}{\left( \alpha .\alpha
\right) }$ is an integer.
\end{enumerate}
It has been remarked that {$\Delta$} provides connections with
matrices and graph theory and can be considered as a strong bridge
between simply  Lie algebras and  modern physics associated with
quivers in string theory compactifications.  This  can be
established  via the so-called Cartan matrix  defined by
\begin{eqnarray} K_{ij}=<\alpha _{i},\alpha _{j}>=2\frac{\alpha
_{i}.\alpha _{j}}{\alpha _{j}.\alpha _{j}}
 \end{eqnarray}
where $\alpha_i$ are simple roots. It has been shown that this
matrix can be encoded in a geometric graph called Dynkin diagram
where the diagonal elements corresponding to vertices and the
non-diagonal ones  are associated with  edges (links)  connecting
the vertices.

A close examination  shows  that one can use such  a connection to
 present a  weighted graph  representation of qubit physics. Indeed, we  explore the connection between Cartan matrices and  Dynkin
diagrams of simply laced Lie algebras to establish such a
correspondence. For such Lie algebras, the Cartan matrices  can be
reduced to
\begin{equation}
K_{ij}=<\alpha_{i},\alpha_{j}>=\alpha_{i}.\alpha_{j}
\end{equation}
where $\alpha_{j}.\alpha_{j}=2$ for all  simple roots $\alpha_{j}$.

Replacing the role played by  the simple roots $\alpha_{i}$ with the
real numbers $a_{i}$  appearing in the density matrix, we can
interpret the density matrix as  a Cartan matrix   using the
following correspondence
\begin{eqnarray} \alpha_{i} &\longrightarrow
& a_{i}  \nonumber\\ \alpha_{i}.\alpha_{j} &\longrightarrow &
a_{i}a_{j}.
\end{eqnarray}
Of course, there are important differences between the two matrices.
Here, though we will  not  be concerned with them. Our objective is
to explore the link   between  graph theory and  simply laced Lie
algebras, via  Dynkin diagrams.

Roughly speaking, the density matrix $\rho$ can be related to the
signless Laplacian matrix $Q(G)$. It is observed that

\begin{equation}
\rho_{ij}=\left\{%
\begin{array}{ll}
    a_i^2D(G)_{ii} &   i=j , \\
   a_{i}a_{j}m_{ij} &    i \neq j \\
\end{array}%
\right.
\end{equation}
 where  $m_{ij}$  are entries of  the adjacency
matrix of the corresponding graph. In this graph, the vertices
$v_{i}$ and $v_{j}$ are connected if $a_{i}a_{j}\neq 0$, otherwise
they are not linked.

Borrowing the idea of Dynkin diagrams of simply laced algebras, we
replace the one-qubit state by a  weighted graph formed by two
vertices $v_{0}$ and $v_{1}$ labeled by their weights $a_{0}^{2}$
and $a_{1}^{2}$,  respectively. These two vertices are linked by an
edge weighted by $a_{0}a_{1}$. Thus, the corresponding graph to this
one-qubit is constructed in the figure 1.

\begin{figure}[tbph] \begin{center}
\includegraphics[width=4cm]{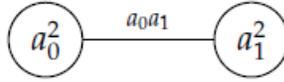}\end{center}
\caption{One-qubit graph representation.}
\end{figure}

This graph can be  considered as a  weighted graph  of two vertices
with a weight function $\omega$ satisfying
\begin{eqnarray}
\omega (v_i,v_j)=\omega (v_j,v_i)=a_ia_j.
\end{eqnarray}
In this way,  the edges correspond to $a_ia_j\neq 0$ and the
non-edges are associated with $a_ia_j=0$. It is recalled that a
simple graph is defined by
\begin{eqnarray}
\omega (v_i,v_j)=0,1 \nonumber\\
\omega (v_i,v_i)=0.
\end{eqnarray}

Note in passing that  signless Laplacian of a  weighted graph can be
considered as a weighted adjacency matrix associated with a quiver
which has two nodes and a single edge carrying  some physical
information. If we forget about the normalization condition,  we can
associate this quiver to a gauge field theory. The vertices $a_i^2$
 correspond to  the  adjoint representation of $\prod_i U(a_i^2)$,
providing a $ G=U(a_0^2) \times U(a_1^2)$ gauge group. The link
$a_0a_1$ is associated   the the Fermi field  bi-fundamental matter.
In type II superstrings, this field theory can be  obtained  when
considering D-branes located the near orbifold points of  the $A_2$
singularity associated with su(3) Lie algebra \cite{18,19}.

\section{Graph theory representation of multi-qubits}

  Having  constructed the   graph associated to one-qubit,
we move now to   the next model associated with  two-qubit quantum
systems defined in a 4 dimensional Hilbert space.  Then, we give the
general statement associated with $n$-qubit systems. In the basis
$\{ |ij>, \; i, j=0,1  \}$, a two-qubit is written as follows
\begin{equation}
|\psi> = a_{00}|00> + a_{01}|01> + a_{10}|10> + a_{11}|11>.
\end{equation}
Up to a  scale factor, the signless Laplacian matrix  can be
identified with  the density matrix. To make contact with  graph
theory,  it is convenient  to consider a  binary index notation
\begin{eqnarray}
p &=& 2^1i+2^0j, \nonumber
\\q &=& 2^1i^{\prime}+2^0j^{\prime},
\end{eqnarray}
where $(i,j,{i}^\prime,j^\prime)=0,1$.   For the two-qubit, the
density matrix can be  written as follows
\begin{eqnarray}
\rho_{pq}=a_{ij}a_{i^\prime j^\prime}.
\end{eqnarray}
This matrix is represented by a graph having four vertices weighted
$a^2_{ij}$ and linked by edges whose weights depend on the values
$a_{ij}a_{i^\prime j^\prime}$. In this way, the two-qubit is
illustrated in the figure 2.
\begin{figure}[tbph] \begin{center}
\includegraphics[width=10cm]{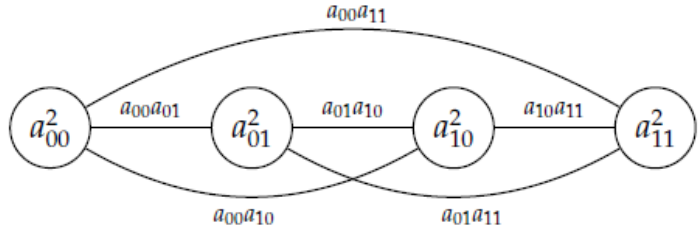}\end{center}
\caption{Two-qubit graph representation.}
\end{figure}

As in the case of one-qubit,  this  graph shares similarities with
quivers  having four vertices and links associated with Fermi field
representations.

 This analysis can be extended to $n$-qubits
associated with $2^n$-dimensional Hilbert spaces. In this way, the
general state reads as
\begin{equation}
\label{qudit} |\psi\rangle=\sum\limits_{i_1\ldots i_n=0,1}a_{
i_1\ldots i_n}|i_1 \ldots i_n\rangle,
\end{equation}
where $a_{ i_1\ldots i_n}$  verify  the  real normalization
condition
\begin{equation}
\label{pcn} \sum\limits_{i_1\ldots i_n=0,1}a_{ i_1\ldots i_n}^2=1.
\end{equation}
 Roughly, the qubit systems can be represented by  a complete weighted graph
 of  $2^n$ vertices. This graph can be obtained from the  corresponding density
 matrix.  As in the two-qubit   example, this matrix   can be written in
 terms of the  binary indices. A close examination shows that  we
 can use the following index notation
\begin{eqnarray}
p&=&\sum_{k=0}^{n-1}2^{i_k}i_k  \nonumber \\
q&=&\sum_{k=0}^{n-1}2^{{i}^\prime_k}{i^\prime}_k
\end{eqnarray}
where $(i_k,i^\prime_k)=0,1$. In  this way,  the  matrix  density
takes the following general form
\begin{eqnarray}
\rho_{pq}=a_{i_1\ldots i_n}a_{i^\prime_1\ldots i^\prime_n},
\end{eqnarray}
which can be associated with the  signless Laplacian matrices  of
  graphs having $2^n$ weighted  vertices and  $2^{n-1}(2^n-1)$
  weighted edges. In this way,  the  weight function $\omega$  can
  be written as
\begin{eqnarray}
\omega (v_p,v_q)=a_{i_1\ldots i_n}a_{i^\prime_1\ldots i^\prime_n}.
\end{eqnarray}

\section{Unitary operations over   graph state  weights}
Having built the graph theory   of qubits, we move now to discuss
the corresponding quantum gates in the graph theory language. We
consider, first, lower dimensional gates, then we propose a general
statement. It is noticed that the classical gates can be obtained by
combining Boolean operations as AND, OR,XOR, NOT and NAND. In fact,
these operations act on classical input bits, taking two values 0
and 1, to produce new bits as output results.  However in  quantum
physics, gates are unitary operators acting  in  $2^n$-dimensional
Hilbert spaces \cite{20,21}. They are considered as simple quantum
circuits. Indeed, the simplest quantum gate performs a unitary
transformation on single-qubit states. Such a quantum gate is called
a single qubit gate. For the  2-dimensional  Hilbert space
representing single-qubit states, the unitary transformation of a
single-qubit gate is given by a $2 \times 2$ unitary matrix.
Examples of single qubit gates  are  the Pauli $X$
and $Z$ gates given,  respectively,    by\\
\begin{eqnarray}
X= \left(%
\begin{array}{cc}
  0 & 1 \\
  1 & 0 \\
\end{array}%
\right), \qquad   Z= \left(%
\begin{array}{cc}
  1 & 0 \\
  0 & -1 \\
\end{array}
\right).
\end{eqnarray}
It is recalled that  $X$ quantum  gate  acts
as follows
\begin{eqnarray}X:\;\left\{%
\begin{array}{ll}
    |0>  \rightarrow |1> &  \\
    |1>  \rightarrow |0>. &  \\
\end{array}%
\right.\end{eqnarray}
 A close examination  shows that
 one-qubit gates  can  be  considered as transformations on the graph
 weights  preserving the  real probability condition,  required by black
 hole applications
 \begin{equation}
 a_0^2+ a_1^2=1.
 \end{equation}
 The  $X$  gate action  can be reduced to the following
 transformation on the weights
 \begin{equation}
 a_i \to  a_{i+1}, \qquad i=0,1.
 \end{equation}
 In the weighted  graph theory,  this   action can be interpreted
 as a permutation operation over the weighted graph associated to one-qubit.
  This  operation permutes the vertex weights and leaves the   edge invariant. This can be illustrated  in
  the figure 3.

\begin{figure}[tbph]
\begin{center} \includegraphics[width=10cm]{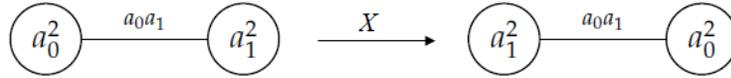}
\caption{X gate action on one-qubit.}
\end{center}
\end{figure}

 However with the  $Z$ gate, the
operation can be understood as
 \begin{equation}
 a_i \to (-1)^i a_{i}, \qquad i=0,1.
 \end{equation}
This  graph operation  is  illustrated   in  the  figure 4.

\begin{figure}[tbph]
\begin{center} \includegraphics[width=10cm]{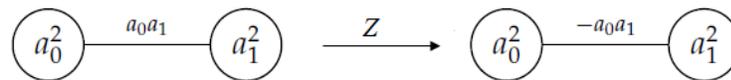}
\caption{Z gate action on one-qubit.}
\end{center}
\end{figure}

As we have seen,  the unitary operations corresponding to the gates
of one-qubit can be reduced to actions over the vertex and edge
weights. We have  seen  how  the  simple quantum gate acts on a
simple qubit, let us see the action over a two-qubit system.  For
these cases, there are  two important universal gates known as
 CNOT and SWAP gates.
In this  graph theory language,    the  CNOT gate  defined by
\begin{eqnarray}
CNOT=\left(
       \begin{array}{cccc}
         1 & 0 & 0 & 0 \\
         0 & 1 & 0 & 0 \\
         0 & 0 & 0 & 1 \\
         0 & 0 & 1 & 0\\
       \end{array}
     \right)
\end{eqnarray}
can be obtained by using the following graph weight actions
\begin{eqnarray}
a_{i\;j} \to a_{i\; i+j}, \qquad i,j=0,1.
\end{eqnarray}
This  operation acts on the corresponding  matrix density  as
follows
\begin{eqnarray}
2^1i+2^0j & \to & 2^1i+2^0(i+j) \nonumber \\ 2^1i^\prime+2^0j^\prime
& \to & 2^1i^\prime+2^0(i^\prime+j^\prime).
\end{eqnarray}
A close inspection shows that the SWAP gate
\begin{eqnarray}
 SWAP=\left(
       \begin{array}{cccc}
         1 & 0 & 0 & 0 \\
         0 & 0  & 1 & 0 \\
         0 & 1 & 0 & 0 \\
         0& 0 & 0 & 1 \\
       \end{array}
     \right)
\end{eqnarray}
can be derived  from the following permutation  action
\begin{eqnarray}
a_{ij} \to a_{ji}, \qquad i=0,1.
\end{eqnarray}

For $n$-qubits,  the   gates are   represented by $2^n\times 2^n$
matrices. In weighted graph theory, they   can be replaced by
actions preserving the real probability condition  (\ref{pcn}). A
simple
 gate can be obtained by the following  index  transformation acting
 on the graph weights $a_{i_1\ldots i_n}$
\begin{eqnarray}
 i_n \to  i_1+\ldots+ i_{n}, \qquad i_k=0,1.
\end{eqnarray}
This  can be considered as a higher dimensional CNOT gate. More
generally, we expect  that this analysis  can be pushed further to
deal with other  non trivial gates to construct non trivial quantum
circuits.

\section{Concluding remarks and speculations  from quiver string theory}
Using the  relation between  simply laced Lie algebras and  Dynkin
diagrams, we have suggested   a new graph theory representation of
quantum states. Using similarities with root systems of such Lie
symmetries,   we   first have discussed one-qubit theory in terms of
a weighted graph having two vertices.  Then, we have  given a
general statement.  In this representation, we have shown that
quantum gates correspond to graph weight actions.

This work comes up  with many open directions and speculations. The
intersecting problem is the discussion of the separability problems
using graph theory methods. This will be addressed elsewhere.
Another  connection   concerns quiver  gauge theories.  As in the
one-qubit, the  weighted graph presented  here  can be considered as
a quiver graph which has two nodes and a single edge carrying  some
physical information correspond to  two factor quiver gauge theory
with Fermi field matter representations. This quiver  could be
obtained by considering D-branes located  on the  $A_2$ like
geometries  associated with su(3) Lie algebra. In this   direction,
the Cartan matrices can be replaced with the density matrix
providing a new way to approach the cohomology class of two
dimensional complex surfaces. This link should be explored to make
contact with quivers associated with   type II superstrings on local
Calabi-Yau manifolds. In fact, we expect that the present weighted
graphs can be explored to build a new class of local
Calabi-manifolds by gluing several non trivial 2-cycles. These
geometries can be motivated from results based on the blowing up  of
ADE singularities used in geometric engineering method of quantum
field theories\cite{22,23}. Similarly as in the ADE cases, one can
build new geometries  using such  a qubit graph theory to deal with
black holes in  type superstrings. We believe this study deserves
more deep reflections. We hope to  come back to this issue in
future.
\newline

\end{document}